\documentclass[onecolumn]{revtex4}
\usepackage{amsmath,amssymb}
\usepackage{amsthm}
\usepackage{epsfig}
\usepackage{xcolor}
\usepackage{fixltx2e}
\usepackage[normalem]{ulem}
\usepackage{siunitx}

\usepackage{xr}
\newcommand{\bsf}{\begin{subfigure}} 
\newcommand{\esf}{\end{subfigure}} 

\graphicspath{{eps/}}            

\begin{document}
\title{Numerical Direct Scattering Transform for Dark Solitons}

\author{Ilya Mullyadzhanov$^{1, 2}$}
\author{Sergey Dremov$^{1, 3}$}
\author{Andrey Gelash$^{1}$}
\email[Corresponding author : ]{A.Gelash@skoltech.ru}

\affiliation{$^{1}$Skolkovo Institute of Science and Technology, Moscow, Russia}
\affiliation{$^{2}$Institute of Automation and Electrometry of SB RAS, Novosibirsk, Russia}
\affiliation{$^{3}$Novosibirsk State University, Novosibirsk, Russia}

\begin{abstract}
We introduce a numerical direct scattering transform scheme for dark solitons of the nonlinear Schrödinger equation, enabling the identification and complete characterization of nonlinear coherent structures in defocusing media with a continuous-wave (CW) background. Our scheme is based on numerically solving the auxiliary Zakharov–Shabat scattering problem with CW boundary conditions and on analytically derived expressions that relate the elements of the transfer matrix to the scattering data for dark solitons and continuous-spectrum waves. To test our approach, we consider two analytically solvable cases of the scattering problem: i) rectangular, and ii) hyperbolic tangent hollows in the CW background, which can contain an arbitrary number of dark solitons, with known scattering data. We revisit the analytical derivations and obtain a complete set of soliton parameters represented by discrete eigenvalues and norming constants. By supplementing the direct scattering transform algorithm with high-precision arithmetic to accurately recover soliton norming constants, we provide a robust method to analyze data from numerical or natural experiments on complex wave fields in optical, hydrodynamical, and other physical systems.
\end{abstract}
\maketitle

\section{Introduction}
Identifying and characterizing nonlinear coherent structures, such as solitons and breathers, embedded in complex wave fields reveals physical principles underlying strongly nonlinear phenomena, e.g., extreme-amplitude wave formation or nonlinear saturation of modulation instability, in systems ranging from optics to the open ocean \cite{osborne1991soliton,conforti2018auto,chekhovskoy2019nonlinear,suret2020nonlinear,redor2019experimental,gelash2019bound,gelash2024bi,lee2026field}. The inverse scattering transform (IST) theory unveils the fundamental significance of the soliton basis in studies of physical systems governed by integrable models such as Korteweg-de Vries or nonlinear Schrödinger equations \cite{NovikovBook1984,AblowitzBook1981}. The IST approach establishes a mathematical connection between soliton parameters, e.g., amplitudes, phases, or positions, and the scattering problem for an auxiliary linear system of ordinary differential equations, which includes the soliton wave field as a potential. With solitons or breathers parameters at hand, one can manage their interactions, predict the characteristic time of Fermi-Pasta-Ulam-Tsingou recurrence, and address the problem of rogue waves formation and statistical properties of spontaneous wave fields in both numerical and natural experiments \cite{xu2019breather,slunyaev2021persistence,gelash2022management,mucci2025manipulation,congy2026exactly}. Other fields of application include prospective nonlinear optical telecommunications \cite{yousefi2014information,turitsyn2017nonlinear,le2017nonlinear} and recently proposed soliton-based remote sensing \cite{dremov2026eigenvalue}, where soliton scattering data retain important information about transmitted signals thanks to the integrable or nearly integrable properties of the propagation channel \cite{NovikovBook1984,AblowitzBook1981,KivsharRMP1989}.

Soliton scattering data is represented by a set of discrete eigenvalues and norming constants $\{\lambda_n,\rho_n\}$, which contain complete information about soliton shape and spatial location. In particular cases, one can obtain this dataset analytically; however, for arbitrary wave fields, this is only possible numerically using an algorithmic version of the direct scattering transform (DST), also known as the direct nonlinear Fourier transform (NFT). The first DST algorithms based on direct integration of the auxiliary scattering problem were proposed in \cite{osborne1990numerical, osborne1991nonlinear, provenzale1991nonlinear, boffetta1992computation, Burtsev1998}. Nowadays, developing efficient DST tools is a broad field of study, with different explicit and implicit numerical schemes used, including Fourier collocations, Töplitz inner bordering, and contour integration \citep{yang2010nonlinear, Frumin2015, wahls2015fast,vasylchenkova2018contour,medvedev2024fast}. Besides the choice of an appropriate numerical scheme, the DST algorithms differ at a more fundamental level in the choice of boundary conditions, which is reflected in the coefficients connecting the elements of the numerically computed transfer matrix \(\bold{\widehat{T}}\) and soliton scattering data. In the case of decaying boundary conditions, these coefficients were derived in \cite{osborne1991nonlinear} and \cite{boffetta1992computation} in the framework of the Korteweg–de Vries (KdV) and the nonlinear Schrödinger equation (NLSE) models.

Another important class of nonlinear coherent structures -- solitons on a continuous-wave (CW) background -- requires asymptotic boundary conditions of constant amplitude. It can be studied using a special version of the IST \cite{zakharov1973interaction,kuznetsov1977,kawata1978,Ma1979}. Recently, we proposed a numerical DST algorithm for solitons of the focusing NLSE on an unstable CW background -- breathers \cite{mullyadzhanov2024numerical}. We now consider the case of the defocusing NLSE (dNLSE) with a stable CW, where coherent structures are dark solitons that emerge as hollows in the background \cite{zakharov1973interaction,kivshar1998dark}, which have been observed experimentally in optics, hydrodynamics, thin magnetic films and Bose-Einstein condensates of cold atoms \cite{weiner1988experimental,chen1993microwave,denschlag2000generating,stellmer2008collisions,chabchoub2013experimental,chabchoub2020phase}. We write the dNLSE in the dimensionless form as
\begin{eqnarray}\label{eqNLSdimentionless}
	i \psi_{t} - \frac12 \psi_{xx} + |\psi|^2 \psi = 0,
\end{eqnarray}
where $\psi(t,x)$ describes a complex-valued wave field. The evolution and spatial variables are time $t$ and coordinate $x$. We impose the following constant-amplitude boundary conditions,
\begin{equation}\label{condensate_boundary_cond}
    \psi \rightarrow A e^{i\Theta_{\pm}} \quad\text{at}\quad x \rightarrow \pm\infty,
\end{equation}
meaning that the wave field asymptotically approaches the background state with generally different phases $\Theta_{\pm}$. As shown in \cite{zakharov1973interaction}, the dNLSE can be integrated using the IST method based on the auxiliary Zakharov-Shabat (ZS) scattering problem. Under boundary conditions (\ref{condensate_boundary_cond}), an arbitrary hollow in the background can contain dark solitons and continuous-spectrum radiation.

Here we develop a numerical DST scheme for the dNLSE, enabling computation of the complete scattering dataset $\{\lambda_n,\rho_n\}$ of dark solitons and the reflection coefficient $r(\lambda)$ of continuous-spectrum waves. We derive analytical expressions that connect the elements of the ZS transfer matrix to the scattering data. Then we compute the transfer matrix using the second-order Boffetta–Osborne scheme \cite{boffetta1992computation} and evaluate soliton parameters. To test our approach, we consider two analytically solvable cases of the ZS problem: i) rectangular, and ii) tanh-shaped hollows in the CW background, which can contain an arbitrary number of dark solitons, with known scattering data. We revisit the analytical derivations \cite{konotop1991randomly,biondini2014spectrum} and obtain a complete set of soliton scattering data $\{\lambda_n,\rho_n\}$. By stabilizing the DST procedure with high-precision arithmetic to recover soliton norming constants, see \cite{mullyadzhanov2019direct,gelash2020anomalous}, we provide a robust way to analyze data from numerical or natural experiments on complex wave fields in defocusing media containing dark solitons, for example, in optics and hydrodynamics, where measurements of the intensity and phase of the wave field are directly accessible \cite{weiner1988experimental,kivshar1998dark,chabchoub2013experimental,chabchoub2020phase}.

\label{Sec:I}

\section{Scattering problem for dark solitons}
\label{Sec:II}
The integration of the NLSE using the IST method is based on the auxiliary linear system introduced by Zakharov and Shabat \cite{zakharov1972exact,zakharov1973interaction}. 
In the defocusing regime, the ZS system takes the form
\begin{eqnarray}\label{ZSsysdefocusing}
	\boldsymbol{\Phi}_{x} &=& \begin{pmatrix} -i \lambda & \psi \\ \psi^* & i \lambda \end{pmatrix}\boldsymbol{\Phi},
	\label{ZSsystem1}\\
	\boldsymbol{\Phi}_t &=& \begin{pmatrix}\ -i\lambda^2 - \frac{i}{2} |\psi|^2 & \lambda \psi + \frac{i}{2} \psi_x \\ \lambda \psi^* - \frac{i}{2} \psi^*_x & i\lambda^2 + \frac{i}{2} |\psi |^2 \end{pmatrix} \boldsymbol{\Phi},
	\label{ZSsystem2}
\end{eqnarray}
where $\boldsymbol{\Phi}(x, t, \lambda) = (\phi_1, \phi_2)^{\mathrm{T}}$ is a two-component vector function, $\lambda$ is a real spectral parameter, and the asterisk denotes complex conjugation. The dNLSE is obtained as the compatibility condition $\boldsymbol{\Phi}_{xt} = \boldsymbol{\Phi}_{tx}$. At a fixed time moment, $\psi(x)$ in Eq.~(\ref{ZSsystem1}) acts as a potential. The DST approach boils down to finding a set of scattering data corresponding to a chosen potential $\psi(x)$ by solving Eq.~(\ref{ZSsystem1}).

To establish scattering data for dark solitons, we consider the asymptotic solution of Eq.~(\ref{ZSsystem1}) as $x \rightarrow \pm\infty$, i.e., when the potential is given by Eq.~(\ref{condensate_boundary_cond}). It can be expressed as a linear combination of two independent vectors
\begin{equation}\label{zs2solfinal}
    \boldsymbol{\Phi} = 
    c_{\text{I}} \begin{pmatrix}\ e^{-\frac{i A^2 t}{2}} \\ p\,e^{-i\Theta_{\pm} + \frac{i A^2 t}{2}} \end{pmatrix}e^{i\lambda\zeta t + i\zeta x} + c_{\text{II}} \begin{pmatrix}\ -p\, e^{i\Theta_{\pm} - \frac{i A^2 t}{2}} \\ e^{\frac{i A^2 t}{2}} \end{pmatrix}e^{-i\lambda\zeta t - i\zeta x},
\end{equation}
where $c_{\text{I}}$ and $c_{\text{II}}$ are arbitrary complex constants, and the functions $\zeta(\lambda)$ and $p(\lambda)$ are defined as
\begin{eqnarray}\label{zeta}
    \zeta &=& \sqrt{\lambda^2 - A^2},
\\\label{rho}
    p &=& \frac{i (\lambda +  \zeta)}{A}.
\end{eqnarray}

Following the quantum mechanical analogy and the standard IST procedure, we introduce the scattering coefficients $a(\lambda)$ and $b(\lambda)$ which fix the asymptotics of the wave function as
\begin{eqnarray}
\label{wave_function}
	\lim_{x\to -\infty}\biggl\{ \boldsymbol{\Phi} &-& \begin{pmatrix}\ 1 \\ p e^{-i\Theta_{-}} \end{pmatrix}e^{i\zeta x}\biggr\} = 0, \label{ScatteringProblemAsymptotics}\\\nonumber
	\lim_{x \to +\infty}\biggl\{\boldsymbol{\Phi} &-& a\cdot\begin{pmatrix}\ 1 \\ p e^{-i\Theta_{+}} \end{pmatrix}e^{i\zeta x} - 
    b\cdot\begin{pmatrix}\ -p e^{i\Theta_{+}} \\ 1 \end{pmatrix}e^{-i\zeta x}\biggr\} = 0.
\end{eqnarray}

To preserve the analyticity of $\zeta(\lambda)$ and general symmetries of operator (\ref{ZSsysdefocusing}), we fix the following sign choice: $\zeta(\lambda) = \operatorname{sign}{(\lambda)} \cdot \sqrt{\lambda^2 - A^2}$ for the continuous spectrum $|\lambda| > A$. For the discrete spectrum $|\lambda| \leq A$, the sign is dictated by the chosen asymptotics; in our case, one has to set $\zeta(\lambda) = -\sqrt{\lambda^2 - A^2}$ for the asymptotics as in (\ref{wave_function}).

According to the IST approach \cite{NovikovBook1984,AblowitzBook1981}, the wave field $\psi(x)$ is in one-to-one correspondence with the scattering data of the scattering problem~(\ref{ZSsystem1}). Similar to quantum mechanics \cite{landau1958quantum}, the full set of scattering data consists of the discrete spectrum $\{\lambda_n, \rho_n\}$ and the continuous spectrum $r(\lambda)$:
\begin{eqnarray}
	&& \big\{
	\lambda_n \,\,|\,\, a(\lambda_n) = 0, \,\, \mathrm{Im}\,\lambda_n = 0
	\big\},\quad \bigg\{\rho_n = \frac{b(\lambda_n)}{a'(\lambda_n)}\bigg\},
    \nonumber\\
	&&
	r(\lambda) = \frac{b(\lambda)}{a(\lambda)}, \qquad
    \lambda\in(-\infty,-A) \cup (A,\infty). \label{ScatteringData}
\end{eqnarray}
In (\ref{ScatteringData}), $a'(\lambda)$ denotes the complex derivative of $a(\lambda)$ with respect to $\lambda$, and $\rho_n$ are the so-called norming constants associated with the eigenvalues $\lambda_n$. The wave field $\psi(x)$ of the dNLSE can be reconstructed from a given set of scattering data (\ref{ScatteringData}) by solving the system of integral Gelfand--Levitan--Marchenko equations \cite{zakharov1973interaction}. For an arbitrary-shaped $\psi(x)$, this reconstruction can only be performed asymptotically at large times and numerically. Scattering data (\ref{ScatteringData}) contain complete information about the physical parameters of solitons and continuous-spectrum waves \cite{NovikovBook1984,AblowitzBook1981}, and its efficient numerical computation is the main task for the DST algorithms.
\section{Analytical solutions for direct scattering problem}
\label{Sec:III}
Numerical DST algorithms can be tested on potentials for which scattering data is known in analytical form. Comprehensive algorithm verification requires complete scattering data information, which is not fully available in the literature. Here we revisit two classical examples: i) rectangular with arbitrary phases, and ii) tanh-shaped hollows in the CW background, which can contain an arbitrary number of dark solitons. The direct scattering problem for these potentials have been considered, e.g., in works \cite{konotop1991randomly,biondini2014spectrum,kang2026deterministic}. Here we accurately obtain the complete scattering data, including soliton norming constants, and present useful details of the derivations below.
\subsection{Inverse step potential}
The inverse step potential allows one to obtain the scattering coefficients in an elementary manner. Consider the wave field given by
\begin{equation}
\label{step_potential}
    \psi(x) = \begin{cases}
        Ae^{i\Theta_-}, \quad\quad x < -\frac{m}{2} \\
        0, \quad\quad\, -\frac{m}{2}< x < \frac{m}{2} \\
        Ae^{i\Theta_+}, \quad\quad\quad x > \frac{m}{2}
    \end{cases},
\end{equation}
where $m$ denotes the characteristic size of the potential inverse step. The scattering problem for this configuration can be solved explicitly \cite{biondini2014spectrum}. The scattering coefficients for potential (\ref{step_potential}) can be written the following form:
\begin{eqnarray}
\label{a_analytics}
a(\lambda) &=& \frac{ e^{-i \zeta m}}{1 + p^2} 
\biggl\{e^{-i\lambda m}
+ p^2 e^{i \lambda m} e^{i(\Theta_+ - \Theta_-)} \biggr\},
\\\label{b_analytics}
b(\lambda) &=& \frac{1}{1+ p^2} 
\biggl\{ -pe^{-i\Theta_{+}} e^{-i \lambda m} + p e^{-i\Theta_{-}} e^{i \lambda m} \biggr\}.
\end{eqnarray}

Hence, we can obtain exprerssion for the coefficient $r(\lambda)$:

\begin{eqnarray}
\label{r_analytics}
r(\lambda) = e^{i\zeta m} \frac{-pe^{-i\Theta_{+}} e^{-i \lambda m} + p e^{-i\Theta_{-}}}{e^{i \lambda m}{e^{-i\lambda m}
+ p^2 e^{i \lambda m} e^{i(\Theta_+ - \Theta_-)}}}.
\end{eqnarray}

In addition we calculate derivative of the scattering coefficient $a(\lambda)$ as follows
\begin{eqnarray}
\label{ad_analytics}
a'(\lambda) &=& \frac{ e^{- i \zeta m}}{1+ p^2} \Biggl[ \left(- i \frac{\lambda}{\zeta} m -\frac{2 p^2}{\zeta (1 + p^2)^2}\right) \left(e^{-i\lambda m}
+ p^2 e^{i \lambda m} e^{i(\Theta_+ - \Theta_-)}\right) +
\\\nonumber
&&\left( -i m e^{-i \lambda m} + \left(i m p^2 + \frac{2 p^2}{\zeta}\right) e^{i \lambda m} e^{i(\Theta_{+} - \Theta_{-})}\right) \Biggr].
\end{eqnarray}

The scattering coefficients derived above provide full access to the scattering data. Specifically, the discrete eigenvalues $\{\lambda_n\}$ can found as solutions of a transcendental equation resulting from the coefficient structure ($a(\lambda_n) = 0$), expressed by the condition:
\begin{eqnarray}
\label{lambda_teq}
e^{-i\lambda_n m}
+ p_n^2 e^{i \lambda_n m} e^{i(\Theta_+ - \Theta_-)} = 0.
\end{eqnarray}
Substituting transcendental condition (\ref{lambda_teq}) into (\ref{b_analytics}) and (\ref{ad_analytics}), and using definition (\ref{ScatteringData}), we obtain analytic expression for soliton norming constants:
\begin{eqnarray}
\label{rhon_analytics}
\rho_n = -\frac{\zeta_n^2e^{i\zeta_n m - i\Theta_+}}{A(m \zeta + 1)}.
\end{eqnarray}
This example provides the simplest nontrivial benchmark for testing the accuracy of the numerical DST algorithm since the scattering coefficients are expressed in closed form through elementary functions. In addition, Eqs.~(\ref{lambda_teq}) and (\ref{rhon_analytics}) represent complete set of discrete scattering data for the inverse step potential (\ref{step_potential}), which can be obtained with the numerical DST tools. Any deviation between the numerically recovered data and the exact expressions allows one to directly quantify the algorithmic errors.

\subsection{Hyperbolic tangent potential}
To validate the DST algorithm's convergence, we need to use smooth potentials, for example, a hyperbolic tangent of arbitrary amplitude $A$ given by
\begin{equation}\label{th_potential}
    \psi(x) = A \cdot \tanh(x).
\end{equation}
In this case system (\ref{ZSsysdefocusing}) acquires the form
\begin{equation}
\label{Phi_tanh}
    	\boldsymbol{\Phi}_{x} = \begin{pmatrix} -i \lambda & A \cdot \tanh(x) \\ A \cdot \tanh(x) & i \lambda \end{pmatrix}\boldsymbol{\Phi}.
\end{equation}
As was shown in \cite{konotop1991randomly}, the ZS scattering problem with the hyperbolic tangent potential can be analytically resolved. Paper \cite{konotop1991randomly} presents a general solution $\boldsymbol{\Phi}$ to system (\ref{Phi_tanh}) and scattering coefficient $a(\lambda)$. Here we revisit the analytical derivation for scattering problem solutuion and obtain a complete set of soliton parameters, including soliton norming constants, see Eq.~(\ref{ScatteringData}).

To solve system (\ref{Phi_tanh}), it is convenient to introduce variables $S = \phi_1 + \phi_2$ and $D = \phi_1 - \phi_2$, which are the sum and difference of the components of the solution vector $\boldsymbol{\Phi}$. Then, from the ZS system, one can obtain two independent second-order differential equations
\begin{eqnarray}\label{ZS_SD_eq}
    S_{xx} + \left[\lambda^2 - A^2 \cdot \tanh^2(x) - A \cdot \mathrm{sech}^2(x) \right]S = 0~, \cr
    D_{xx} + \left[\lambda^2 - A^2 \cdot \tanh^2(x) + A \cdot \mathrm{sech}^2(x) \right]D = 0~.
\end{eqnarray}
In terms of variable $z = (1 - \tanh(x))/2$ system (\ref{ZS_SD_eq}) takes the following form:
\begin{eqnarray}\label{ZS_SD_z_eq}
    z(z-1) S''_{zz} + (2z - 1) S'_z + \left[\frac{\lambda^2 - A^2}{4z(z-1)} - A(A-1) \right]S = 0~,  \cr
    z(z-1) D''_{zz} + (2z - 1) D'_z + \left[\frac{\lambda^2 - A^2}{4z(z-1)} - A(A+1) \right]S = 0~,
\end{eqnarray}
which has three regular singular points: $z = 0, z =1, z = \infty$, and therefore suggests hypergeometric functions as potential solutions, similar to the case of hyperbolic secant potential in the focusing NLSE considered in \cite{satsuma1974b}. Indeed, by taking $S(z)$ and $D(z)$ in the form $z^a \cdot (1-z)^b \cdot \omega(z)$ (Frobenius method), one obtains hypergeometric equation for $\omega(z)$ with $a = b = i\zeta/2$. Finally, applying asymptotic conditions (\ref{ScatteringProblemAsymptotics}) results in:
\begin{eqnarray}\label{ZS_dNLS_SD_solutions}
    S(z) =  (1-p) \cdot z^{i\zeta/2}(1-z)^{i\zeta/2} \cdot {}_2F_1(i\zeta + A,i\zeta - A + 1,i\zeta + 1,1-z)~,  \cr
    D(z) =  (1+p) \cdot z^{i\zeta/2}(1-z)^{i\zeta/2} \cdot {}_2F_1(i\zeta - A,i\zeta + A + 1,i\zeta + 1,1-z)~.
\end{eqnarray}

Note that hypergeometric functions ${}_2F_1$ are taken with argument $1-z$ (i.e., around $z = 1$ or $x = -\infty$). This choice is convenient to satisfy the boundary conditions for the right scattering problem, exactly as in (\ref{ScatteringProblemAsymptotics}).

To find scattering coefficients $a(\lambda)$ and $b(\lambda)$ it is not even necessary to know the exact expressions for $\phi_1$ and $\phi_2$. It is only necessary to construct boundary conditions for $S(z)$ or $D(z)$. Therefore, we express $S(z)$ from (\ref{ZS_dNLS_SD_solutions}) around $z = 0 ~(x = \infty)$ by using $z \approx e^{-2x}$ and connection formulas for hypergeometric functions:
\begin{equation}\label{S_solution_def_inf}
    S(z\rightarrow 0) =  (1-p) \cdot e^{-i\kappa x} \frac{\Gamma(i\kappa + 1)\Gamma(-i\kappa)}{\Gamma(1 - A) \Gamma(A)} + (1-p) \cdot e^{i\kappa x} \frac{\Gamma(i\kappa + 1)\Gamma(i\kappa)}{\Gamma(i\kappa + A) \Gamma(i\kappa - A + 1)}~.
\end{equation}

Here $\Gamma(x)$ is the Euler's Gamma function. Thus, the scattering coefficients are:
\begin{eqnarray}\label{coeffs_a}
    a(\lambda) &=& \frac{(1 - p)}{(1 + p)}\frac{\Gamma(i\zeta + 1)\Gamma(i\zeta)}{\Gamma(i\zeta + A) \Gamma(i\zeta - A + 1)} = -\frac{\zeta}{\lambda}\frac{\Gamma^2(i\zeta)}{\Gamma(i\zeta + A) \Gamma(i\zeta - A)}, 
    \\
    \label{coeffs_b}
    b(\lambda) &=& \frac{\Gamma(i\zeta + 1)\Gamma(-i\zeta)}{\Gamma(1 - A) \Gamma(A)} = \frac{i \cdot \sin(\pi A)}{\sinh(\pi\zeta)}~.
\end{eqnarray}

Then the continuous spectrum of the hyperbolic tangent potential can be expressed as
\begin{eqnarray}\label{coeff_r}
    r(\lambda) = -\frac{i \cdot \sin(\pi A)}{\sinh(\pi\zeta)}\frac{\lambda}{\zeta}\frac{\Gamma(i\zeta + A) \Gamma(i\zeta - A)}{\Gamma^2(i\zeta)}, \quad \lambda\in(-\infty,-A) \cup (A,\infty).
\end{eqnarray}

One can easily check that for the continuous spectrum, the obtained scattering coefficients satisfy relation $|a(\lambda)|^2 - |b(\lambda)|^2 = 1$. The discrete spectrum $\{\lambda_n \}$ is determined by the zeros of scattering coefficient $a(\lambda)$, which correspond to simple poles of the Gamma functions in the denominator of expression (\ref{coeffs_a}), i.e. when:
\begin{eqnarray}\label{DS_cond}
i\zeta_n + A &=& -n,  \cr
i\zeta_n - A &=& -n,
\end{eqnarray}
where $\zeta_n = \zeta(\lambda_n)$, leading to expression for discrete spectrum eigenvalues of the hyperbolic tangent potential:
\begin{equation}
\label{tanh_eigs}
    \lambda_n = \pm\sqrt{n(2A - n)}, \qquad 0 \le n < A .
\end{equation}
To the best of our knowledge, the discrete spectrum expression (\ref{tanh_eigs}) was first obtained in \cite{zhao1989propagation}, and later derived in \cite{konotop1991randomly} and studied in \cite{kivshar1993dark}.

According to (\ref{tanh_eigs}), there is always soliton with $\lambda_0 = 0$ in hyperbolic tangent potential. When $A>1/2$ potential (\ref{th_potential}) also contains $n$ pairs of solitons with $\lambda_n = \pm\sqrt{n(2A - n)}$ where $n < A + 1/2$. To obtain norming constants $\rho_n$, we first calculate scattering coefficient $b(\lambda)$ at discrete eigenvalue points (\ref{tanh_eigs}) according to definition (\ref{ScatteringData}):
\begin{equation}
\label{bn}
    b(\lambda_n) = (-1)^{n+1}\,,
\end{equation}
and then we evaluate the derivative of $a(\lambda)$ via digamma functions $\Psi(x)$:
\begin{equation}\label{da_final}
    \frac{da}{d\lambda} = a(\lambda) \left[\frac{A^2}{\lambda \zeta^2}  +\frac{i\lambda}{\zeta}(2 \Psi(i\zeta) - \Psi(i\zeta + A) - \Psi(i\zeta - A)) \right].
\end{equation}
In the limit $\lambda \to \lambda_n$ one needs to use the expression $\lim\limits_{z \to -n} \frac{\Psi(z)}{\Gamma(z)} = (-1)^{n+1} n!$ and individually consider the case $\lambda_0 = 0$, which altogether yields:
\begin{equation}
\label{dan}
    a'(\lambda)\big\vert_{\lambda=\lambda_n} = 
    \begin{cases}
        -i \cdot (-1)^n \cdot n! \cdot\frac{\Gamma^2(A-n)}{\Gamma(2A-n)}, \quad\quad\,\,\, n\ne 0 \\
        -\frac{i \cdot \Gamma^2(A)}{2 \cdot \Gamma(2A)}, \quad\quad\quad\quad\quad\quad\quad\quad\quad n=0
    \end{cases},
\end{equation}
Finally using definitions (\ref{bn},\ref{dan}) and Kroneker's delta $\delta_{mn}$, we obtain expression for the norming constants:
\begin{equation}\label{rho_final}
    \rho_{n} = \frac{-i (1 + \delta_{n0}) \cdot\Gamma(2A - n)}{n! \cdot\Gamma^2(A-n)}~.
\end{equation}
Expressions (\ref{coeff_r}), (\ref{tanh_eigs}) and (\ref{rho_final}) present a complete set of scattering data for hyperbolic tangent potential (\ref{th_potential}). 

\section{Direct scattering transform algorithm}
\label{Sec:IV}
This section outlines our algorithmic implementation of the direct scattering transform (DST) for the dNSLE with boundary conditions (\ref{condensate_boundary_cond}). The objective is to compute the full scattering data set (\ref{ScatteringData}) numerically. Our method extends the Boffetta–Osborne algorithm, originally developed for the zero-background case \cite{boffetta1992computation}.

We restrict the DST problem to a finite computational domain \(x \in [-L/2, L/2]\) instead of the infinite line. The boundary conditions (\ref{wave_function}) are shifted from \(x \to \pm\infty\) to \(x = \pm L/2\). Similar to our previous works \cite{gelash2020anomalous,mullyadzhanov2021magnus}, we define a truncated wave function \(\boldsymbol{\Phi}_{\mathrm{tr}}\) along with the truncated scattering coefficients \(a_{\mathrm{tr}}(\lambda)\) and \(b_{\mathrm{tr}}(\lambda)\), which satisfy the shifted boundary conditions:
\begin{eqnarray}
\label{ScatteringProblemTrunct}
\boldsymbol{\Phi}_{\mathrm{tr}}(-L/2) &=& 
\begin{pmatrix} 1 \\ p e^{-i\Theta_{-}} \end{pmatrix} e^{-i\zeta L/2},
\\\nonumber
\boldsymbol{\Phi}_{\mathrm{tr}}(L/2) &=& 
a_{\mathrm{tr}} \begin{pmatrix} 1 \\ p e^{-i\Theta_{+}} \end{pmatrix} e^{i\zeta L/2} + 
b_{\mathrm{tr}} \begin{pmatrix} -p e^{i\Theta_{+}} \\ 1 \end{pmatrix} e^{-i\zeta L/2}.
\end{eqnarray}

Following \cite{boffetta1992computation}, we form a four-component vector \((\boldsymbol{\Phi}_{\mathrm{tr}}, \boldsymbol{\Phi}'_{\mathrm{tr}})^{\mathrm{T}}\) and introduce a \(4\times 4\) transfer matrix \(\bold{\widehat{T}}\), that links the solution at the left and right boundaries:
\begin{equation}
\label{T_matrix}
\begin{pmatrix}
\boldsymbol{\Phi}_{\mathrm{tr}} (L/2,\lambda) \\
\boldsymbol{\Phi}'_{\mathrm{tr}} (L/2,\lambda)
\end{pmatrix}
=
\underbrace{
\begin{pmatrix}
\boldsymbol{\widehat{\Sigma}} & 0\\
\boldsymbol{\widehat{\Sigma}}' & \boldsymbol{\widehat{\Sigma}}
\end{pmatrix}
}_{\bold{\widehat{T}}}
\begin{pmatrix}
\boldsymbol{\Phi}_{\mathrm{tr}} (-L/2,\lambda) \\
\boldsymbol{\Phi'}_{\mathrm{tr}} (-L/2,\lambda)
\end{pmatrix},
\end{equation}
where \(\boldsymbol{\widehat{\Sigma}}(\lambda)\) is the \(2\times 2\) transfer matrix for \(\boldsymbol{\Phi}_{\mathrm{tr}}\), i.e., \(\boldsymbol{\Phi}_{\mathrm{tr}}(L/2) = \boldsymbol{\widehat{\Sigma}} \, \boldsymbol{\Phi}_{\mathrm{tr}}(-L/2)\).

We denote the center of the \(m\)-th bin as \(x_m\) with width \(\Delta x_m\) and $x_c = x_m - \Delta x_m / 2$, $x_p = x_m + \Delta x_m / 2$. The evolution of the wave function across a bin is given by
\begin{eqnarray}
\label{MagnusRel} \boldsymbol{\Phi}(x_p) = \widehat{\boldsymbol{U}}(x_m) \boldsymbol{\Phi}(x_c), 
\end{eqnarray}
which allows us to express the matrix \(\bold{\widehat{\Sigma}}\) as the ordered product
\begin{eqnarray}
\label{SigmaMat} \bold{\widehat{\Sigma}} = \prod_{m=1}^{M} \widehat{\boldsymbol{U}} (x_m),
\end{eqnarray}
thereby providing a means to compute \(\bold{\widehat{T}}\) according to Eq.~(\ref{T_matrix}).

The transfer matrix for a single bin takes the form
\begin{eqnarray}
\widehat{\boldsymbol{U}}(x_m) =
\begin{pmatrix}
\cosh k_m - \frac{i \varkappa_m}{k_m} \sinh k_m & \frac{\chi_m}{k_m} \sinh k_m \\ 
\frac{- \chi^*_m}{k_m} \sinh k_m & \cosh k_m + \frac{i \varkappa_m}{k_m} \sinh k_m
\end{pmatrix},
\label{eqMexp1}
\end{eqnarray}
which corresponds to the Boffetta–Osborne second-order scheme with the following parameters:
\begin{eqnarray}
            && k_m^2 = (- |\psi_m|^2 - \lambda^2) \Delta x_m^2,      
 \\
            && i \varkappa_m = i \lambda \Delta x_m, \\
            && \chi_m = \psi_m \Delta x_m.
\end{eqnarray}

The matrix of \(\lambda\)-derivatives \(\widehat{\boldsymbol{U}}'(x_m) = \partial_\lambda \widehat{\boldsymbol{U}}\), which determines \(\bold{\widehat{\Sigma}}'\), has the following elements \(U'_{ij}\) (with \(i,j=1,2\)):
\begin{eqnarray}
&& U'_{11} = \left( k_m' - \frac{i \varkappa_m'}{k_m} + \frac{i \varkappa_m k_m'}{k_m^2} \right) \sinh k_m - \frac{i \varkappa_m k_m'}{k_m} \cosh k_m, \\
&& U'_{12} = \left( \frac{\chi_m'}{k_m} - \frac{\chi_m k_m'}{k_m^2} \right) \sinh k_m + \frac{\chi_m k_m'}{k_m} \cosh k_m, \\
&& U'_{21} = - \left( \frac{\chi_m'^*}{k_m} - \frac{\chi_m^* k_m'}{k_m^2} \right) \sinh k_m - \frac{\chi_m^* k_m'}{k_m} \cosh k_m, \\
&& U'_{22} = \left( k_m' + \frac{i \varkappa_m'}{k_m} - \frac{i \varkappa_m k_m'}{k_m^2} \right) \sinh k_m + \frac{i \varkappa_m k_m'}{k_m} \cosh k_m,
\label{eq4orUzeta}
\end{eqnarray}
where the derivatives \(k_m' = \partial_\lambda k_m\), \(\chi_m' = \partial_\lambda \chi_m\), \(\varkappa_m' = \partial_\lambda \varkappa_m\) are obtained from the definitions above:
\begin{eqnarray}
\label{km2dzeta}  && k_m' = \frac{1}{k_m} (-\lambda \Delta x_m^2), \\
\label{zmdzeta}   && \varkappa_m' = \Delta x_m, \\
\label{chimdzeta} && \chi_m' = 0.
\end{eqnarray}

The computation of the transfer matrix is universal, i.e. it does not depend on the boundary conditions, so that expressions provided above repeat the Boffetta–Osborne algorithm. To make the DST algorithm work with dark solitons, here, using Eq.~(\ref{ScatteringProblemTrunct}) and Eq.~(\ref{T_matrix}) we derive the following connection between the truncated scattering coefficients and the elements $T_{ij}$ of the matrix $\bold{\widehat{T}}$:
\begin{eqnarray}
\label{a2T}
a_{\mathrm{tr}}(\lambda) &=& e^{-i L \zeta} \frac{T_{11} + T_{12} p e^{-i \Theta_-} + T_{21} p e^{i \Theta_+} + T_{22} p^2 e^{i(\Theta_+ - \Theta_-)}}{1 + p^2},
\\
\label{b2T}
b_{\mathrm{tr}}(\lambda) &=& \frac{-T_{11} p e^{-i \Theta_+} - T_{12} p^2 e^{-i \Theta_-} e^{-i \Theta_+} + T_{21} + T_{22} p e^{-i \Theta_-}}{1 + p^2},
\end{eqnarray}
and analogous connection for the derivative of the transmission coefficient:
\begin{eqnarray}
\label{da2T}
a'_{\mathrm{tr}}(\lambda) = e^{-i L \zeta} \frac{\left(-\frac{2 p^2}{(1 + p^2) \zeta} - \frac{i L \lambda}{\zeta}\right)\left(T_{11} + T_{12} p e^{-i \Theta_-} + T_{21} p e^{i \Theta_+} + T_{22} p^2 e^{i(\Theta_+ - \Theta_-)}\right)}{1 + p^2} +
\\\nonumber
\frac{T_{31} + T_{32} p e^{-i \Theta_-} + T_{12} \frac{p}{\zeta} e^{-i \Theta_-} + T_{41} p e^{i \Theta_+} + T_{21} \frac{p}{\zeta} e^{i \Theta_+} + T_{21} \frac{p}{\zeta} + 2 \frac{p}{\zeta} p e^{i \Theta_+} e^{-i \Theta_-} T_{22} + p^2 e^{i \Theta_+} e^{i (\Theta_+ - \Theta_-)} T_{42}}{1 + p^2}.
\end{eqnarray}

The expressions presented in Eqs.~(\ref{a2T}-\ref{da2T}) are the key feature of the proposed DST for dark solitons of the dNLSE model with the wave field asymptotics provided by Eq.~(\ref{condensate_boundary_cond}). Together with our previous work on DST for breathers in the fNLSE model \cite{mullyadzhanov2024numerical}, we complete the extension of the Boffetta–Osborne algorithm to the case of continuous wave boundary conditions.

\section{Numerical examples}
\label{Sec:V}

\begin{figure}[!t]
	\centering
	\includegraphics[width=1.05\linewidth]{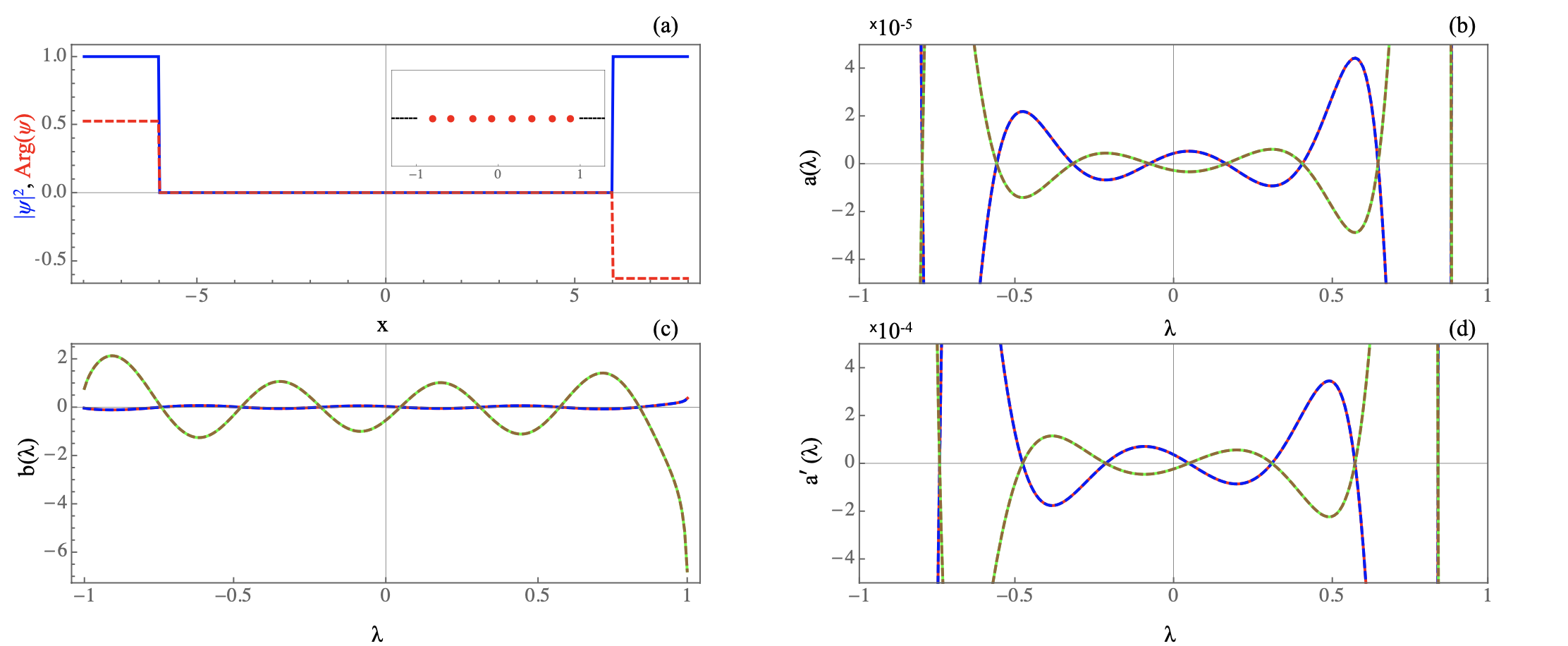}
	\caption{\small Verification of the developed DST algorithm using inverse step potential: (a) The wave field is represented by a potential that vanishes on $[-m/2,m/2]$ with $m=12$ (blue line); a red dashed line shows phase behavior. The inset shows eigenvalues as red dots, and the branch cut of the function $\zeta(\lambda)$, see Eq.~(\ref{zeta}), is indicated by a black dotted line. (b), (c), (d) present the behavior of the scattering coefficients $a(\lambda)$, $b(\lambda)$, and $a'(\lambda)$, respectively. Numerical results (solid lines) are compared with analytical results (dashed lines) to validate the algorithm. Blue and red colors correspond to the real part, while green and brown correspond to the imaginary part. Panel (b) shows where the real and imaginary parts of $a(\lambda)$ simultaneously cross zero, which highlights the location of the eigenvalues, as the eigenvalues follow from the zeros of $a(\lambda)$.}
	\label{fig:1}
\end{figure}

To verify the proposed algorithm, we consider three examples of the dNLSE wave fields containing dark solitons: i) inverse step and hyperbolic tangent potential
for which scattering coefficients and scattering data is known analytically, see section~\ref{Sec:III}, and ii) a randomly modulated hollows in the CW background for which exact solution is not available, however convergence test can confirm that our algorithm performs correct.

Our DST algorithm, implemented in Wolfram Mathematica, works as follows. It receives a discretized wave field (potential) and identifies its asymptotic phases $\Theta_{\pm}$ to specify it in the connection relations (\ref{a2T}-\ref{da2T}). Then, on a rough discrete grid with step size $\Delta\lambda\sim 10^{-2}$ over the interval $[-A,A]$, we compute the transfer matrix and, subsequently, the scattering coefficients using the algorithm described in Section IV. We identify a set of zeros of the transmission coefficients with an accuracy $\sim\Delta\lambda$ and then use them as seed points for the Newton-Raphson method to find a set of discrete eigenvalues $\{\lambda^{\text{num}}_n\}$ with high precision. Then, we take scattering coefficients evaluated at the found discrete eigenvalues and compute soliton norming constants $\{\rho^{\text{num}}_n\}$. In addition, we compute scattering coefficients on a discrete grid $\{ \lambda_i \}$ over the interval $[A_{\text{min}},-A]\cup [A,A_{\text{max}}]$ and evaluate the reflection coefficient $r^{\text{num}}(\lambda_i)$.

\begin{figure}[!t]
	\centering
	\includegraphics[width=1.1\linewidth]{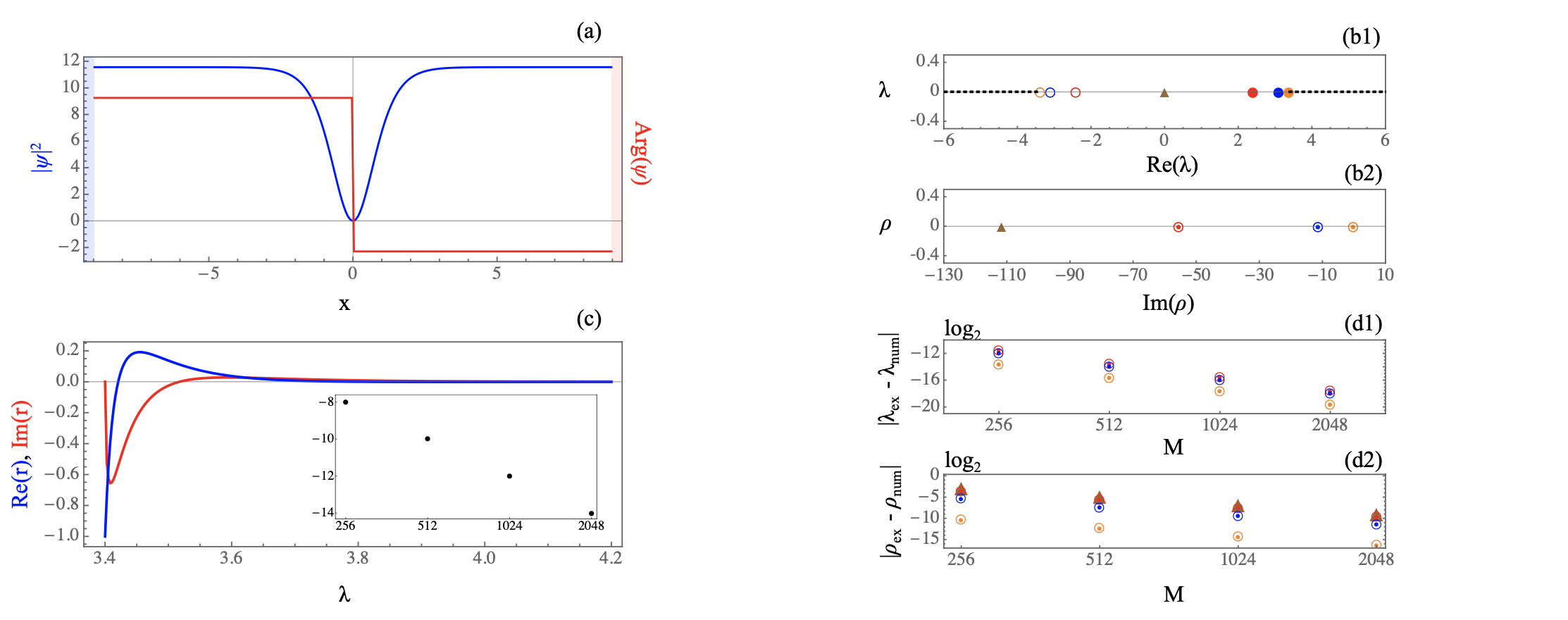}
	\caption{\small 
    Verification of the developed DST algorithm using hyperbolic tangent potential. 
    Panel (a) shows the wave field (blue line) represented by $A \tanh(x)$ with $A = 3.4$, and the red line shows the phase behavior. 
    Panel (b1) illustrates the $\lambda$-plane of the spectral parameter: 
    the branch cut of the function $\zeta(\lambda)$, see Eq.~(\ref{zeta}), is shown as a black dotted line, and the phase constants $\rho_n$ are displayed in panel (b2). 
    Filled and empty circles denote positive and negative eigenvalues $\lambda_n$ (which are real) 
    and the corresponding phase constants $\rho_n$ (which are imaginary), respectively. 
    Panel (c) shows the behaviour of the real (blue) and imaginary (red) parts of reflection coefficient $r(\lambda)$ representing the continuous spectrum, where $\lambda \in (-\infty, -A) \cup (A, \infty)$; since the coefficient has a $\lambda$-symmetry, $(A, \infty)$ branch is shown only. The inset displays the convergence $\log_2|r_{\text{ex}} - r_{\text{num}}|$ with the discretisation parameter on the $x$-axis; panels (d1) and (d2) show second-order convergence for $\lambda_n$ and $\rho_n$, respectively; $M$ denotes the discretisation parameter.
	}
	\label{fig:2}
\end{figure}

Fig.~\ref{fig:1} shows results of our first numerical test made using an inverse step potential. At the chosen parameters, the potential has eight discrete eigenvalues located inside the interval $[-A,A]$, see Fig.~\ref{fig:1}(a). Asymptotic phases of the inverse step are chosen differently in order to provide a generally fair verification. In this test, using our DST algorithm, we compute the scattering coefficients and the derivative of the transmission coefficient, and compare analytical expressions (\ref{a_analytics}), (\ref{b_analytics}), and (\ref{ad_analytics}) with numerical results, which demonstrate an accurate match, see Fig.~\ref{fig:1}(b,c,d). Note, that locations of zeros of the transmission coefficient $a(\lambda)$, see Fig.~\ref{fig:1}(b) corresponds to analytical discrete eigenvalues provided by Eq.~(\ref{lambda_teq}).

We perform a second test using a smooth hyperbolic tangent potential (\ref{th_potential}) with $A=3.4$, which has seven discrete eigenvalues, see Fig.~(\ref{fig:2}). Due to potential symmetry, discrete eigenvalues are located symmetrically, and the norming constants of symmetric eigenvalues coincide, see Fig.~(\ref{fig:2})(b1,b2). We run our DST algorithm and identify the complete set of scattering data on different grids with $M = 2^K$ discretization points, which allows us to perform the algorithm convergence test. As expected, when compared with analytical expressions (\ref{tanh_eigs}), (\ref{coeff_r}) and (\ref{rho_final}), our DST algorithm demonstrates second-order convergence for discrete eigenvalues, norming constants and continuous spectrum, see Fig.~(\ref{fig:2})(b,d). Note that the zero eigenvalue $\lambda_0 = 0$ of the hyperbolic tangent potential is a special case with respect to the DST algorithm convergence. When searching for the zero of the transmission coefficient at $\lambda=\lambda_0$, we find it with very high accuracy $|\lambda_0-\lambda^{\text{num}}_0| \lesssim  10^{-30}$ even at rough potential discretizations. Further potential detailization does not change this initially tiny computational error; i.e., no convergence is present. This feature is explained by the structure of transfer matrix product (\ref{SigmaMat}) at $\lambda=\lambda_0$, which reduces to the product of only two matrices:
\begin{eqnarray}
\mathbf{\hat U}(x_m) &=& \begin{pmatrix}
\cosh(k_m) & -\sinh(k_m) \\
-\sinh(k_m) & \cosh(k_m)
\end{pmatrix}, \qquad x_m < 0\,,
\\
\mathbf{\hat U}(x_m) &=& \begin{pmatrix}
\cosh(k_m) & \sinh(k_m) \\
\sinh(k_m) & \cosh(k_m)
\end{pmatrix}, \qquad\quad\,\,\,\,\, x_m > 0\,,
\\
\mathbf{\hat U}(x_m) &=& \begin{pmatrix}
1 & 0 \\
0 & 1
\end{pmatrix}, \qquad\qquad\qquad\qquad\qquad\,\,\,\,\, x_m = 0\,.
\end{eqnarray}
Their product (\ref{SigmaMat}) ultimately yields the identity matrix $\mathbf{\hat E}$ as one can deduce from the pairwise property $\mathbf{\hat U}(-x_m)\cdot\mathbf{\hat U}(x_m) = \mathbf{\hat E}$, which implies no dependence on the discretization parameter and the absence of convergence of computational errors. At the same time, computing the norming constant $\rho_0$ again yields second-order convergence; see Fig.~\ref{fig:2}(d2).

\begin{figure}[!t]
	\centering
	\includegraphics[width=1.0\linewidth]{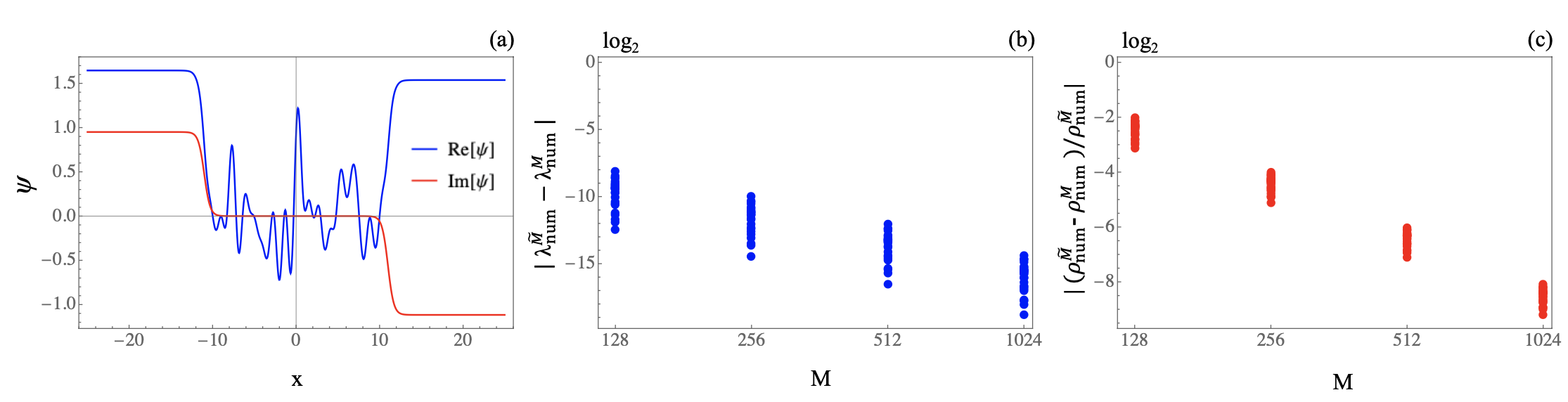}
	\caption{\small Verification of the developed DST algorithm using an arbitrary-modulated hollow in the CW background as a test potential. (a) Real and imaginary parts of the potential modeled by Eq.~(\ref{arb_potential}). (b,c) Second-order convergence of numerical errors for the identified $N=28$ dark solitons, evaluated at different number of discretization points $M$ with respect to analogous data obtained for the most accurate potential detalization at $\Tilde{M} = 2^{11}$.}
	\label{fig:3}
\end{figure}

Finally, we test the performance of our DST algorithm using an arbitrary-modulated hollow in the CW background, modeled by the following expression
\begin{eqnarray}
\label{arb_potential}
\psi(x) = \frac{A e^{i\Theta_-}}{2} \left(1 - \tanh\left(\frac{x + a_r/2}{\sigma_1}\right)\right)
+ \frac{A e^{i\Theta_+}}{2} \left(1 + \tanh\left(\frac{x - a_r/2}{\sigma_1}\right)\right)
+ 
\\\nonumber
B e^{-(x/\sigma_2)^8} \sum_{j=1}^{10} \Bigl[ a_j \sin(k_j x) + b_j \cos(k_j x) \Bigr].
\end{eqnarray}

The potential (\ref{arb_potential}) is defined as a superposition of two smooth step-like barriers and an oscillatory addition. The parameters $A$, $\Theta_-$ and $\Theta_+$ determine the amplitude of the background (step-like) potential and sets the height and phases of the plateaus according to boundary conditions Eq.~(\ref{condensate_boundary_cond}). The parameter $a_r$ sets the distance between the centers of the two step-like transitions. The steepness of the transitions is controlled by the parameter $\sigma_1$. The oscillatory addition represents a sum of 20 harmonics, each being a linear combination of sine and cosine with its own coefficients $a_j$ and $b_j$ and wavenumbers $k_j$. All coefficients $a_j$ and $b_j$ were chosen randomly in the range from $0$ to $5$. The wavenumbers $k_j$ are also random variables. The amplitude of the entire oscillatory addition is controlled by the parameter $B$. To localize the fluctuations in the central region and suppress them at the edges, we use the factor $\exp[-(x/\sigma_2)^8]$, where the parameter $\sigma_2$ sets the characteristic decay scale.

Exact scattering data for potential (\ref{arb_potential}) is not available and we study convergence of the numerical results at different number of discretization points $M$ with respect to analogous data obtained for the most accurate potential detalization at $\Tilde{M} = 2^{11}$. Fig.~(\ref{fig:3}) shows an example of an arbitrary-modulated hollow with different asymptotic phases and numerical errors for the identified $N=28$ dark solitons. For such a large number of discrete eigenvalues, computation of norming constants suffers from anomalous errors of the DST \cite{gelash2020anomalous}. To avoid the anomalous errors and also the round-off errors of the calculating the transfer matrix we employ high-precision arithmetic operations, as suggested in \cite{mullyadzhanov2019direct,gelash2020anomalous}. Numerical data demonstrates second order convergence of the errors evaluated as $|\lambda_{\text{num}}^{M} - \lambda_{\text{num}}^{\Tilde{M}}|$ for soliton eigenvalues and $|(\rho_{\text{num}}^{M} - \rho_{\text{num}}^{\Tilde{M}})/(\rho_{\text{num}}^{\Tilde{M}})|$ for soliton norming constants.

\section{Conclusions}
Our work generalizes the classical Boffetta–Osborne DST algorithm \cite{boffetta1992computation} to the case of dark solitons in the dNLSE model. Together with our previous work on the DST for breathers of the fNLSE \cite{mullyadzhanov2024numerical}, we complete the family of DST algorithms under continuous-wave boundary conditions. The main feature of our approach is the derived expressions (\ref{a2T}), (\ref{b2T}) and (\ref{da2T}) that connect the truncated scattering coefficients \(a_{\mathrm{tr}}(\lambda)\), \(b_{\mathrm{tr}}(\lambda)\) and \(a'_{\mathrm{tr}}(\lambda)\) to the elements of the $4\times 4$ transfer matrix $\bold{\widehat{T}}$, that being consistent with the asymptotic definition of the scattering data (\ref{wave_function}) provides access to continuous spectrum reflection coefficient $r(\lambda)$ and dark soliton eigenvalues and norming constants $\{\lambda_n, \rho_n\}$ of an arbitrary-shaped wave fields of the dNLSE model under the boundary conditions (\ref{condensate_boundary_cond}). Note that under rapidly decaying boundary conditions, the dNLSE model does not support solitons, and the continuous spectrum reflection coefficient can be evaluated using the standard Boffetta–Osborne approach or alternative algorithms such as Töplitz inner bordering \cite{frumin2015efficient}. Also, in the case of the same asymptotic phases $\Theta_{+}=\Theta_{-}$, dark soliton discrete eigenvalues can be computed using the Fourier collocation method \cite{yang2010nonlinear}. Under periodic boundary conditions, the scattering data exhibit a fundamentally different finite-gap structure \cite{NovikovBook1984,matveev200830} and, e.g., for the dNLSE, can be evaluated using the approach proposed by A.~Osborne in \cite{osborne1993numerical}, see also recently proposed approaches for another integrable models \cite{bilman2022computation}.

To provide a comprehensive verification of our DST algorithm, we revisited the analytical solution of the ZS problem for the inverse-step and hyperbolic-tangent potentials. Compared to previous works \cite{konotop1991randomly,biondini2014spectrum}, here we complete the scattering data set by including the norming constants $\{\rho_n\}$ for dark solitons, see Eqs.~(\ref{rhon_analytics}) and (\ref{rho_final}). Our numerical tests demonstrate accurate computation of the scattering coefficients and the complete set of scattering data for potentials containing $N=8$ and $N=7$ dark solitons, see Figs.~(\ref{fig:1}) and (\ref{fig:2}). In our final test, we identify $N=28$ dark solitons in a randomly modulated hollow of the continuous-wave background; see Fig.~(\ref{fig:3}). Following our previous works \cite{mullyadzhanov2019direct,gelash2020anomalous,agafontsev2023bound}, we employ high-precision arithmetic operations to find soliton eigenvalues with 100-digit precision, which is necessary to overcome anomalous errors in the evaluation of their norming constants. In addition, we confirm second-order convergence of our algorithm by performing numerical computations on different numerical grids as illustrated by Figs.~(\ref{fig:2}) and (\ref{fig:3}). Further developments can be devoted to applications of high order convergence schemes and demonstrations of the DST performance using large ensembles of dark multi-soliton complexes \cite{mullyadzhanov2019direct,medvedev2019exponential,mullyadzhanov2021magnus}. As a nonlinear generalization of Fourier decomposition, the DST has been used to analyze nonlinear wave fields in various physical systems \cite{AblowitzBook1981,OsborneBook2010,turitsyn2017nonlinear,sugavanam2019analysis}. The developed robust DST algorithm for dark solitons provide a versatile way to analyze data from numerical or natural experiments on complex wave fields where measurements of the intensity and phase of the wave field are directly accessible, for example, in optics and hydrodynamics \cite{weiner1988experimental,kivshar1998dark,chabchoub2013experimental,chabchoub2020phase}, including the emerging field of theoretical and experimental studies on strongly interacting soliton gases and integrable turbulence \cite{zakharov2009turbulence,agafontsev2015integrable,kraych2019statistical,shurgalina2016nonlinear,gelash2019bound,suret2024soliton,congy2026exactly}. Another promising direction is the DST analysis of nonlinear coherent structures in nonitegrable models close to the dNSLE, such as driven dissipative optical microresonators and exciton-polariton condensates \cite{lobanov2015frequency,xue2016normal,smirnov2014dynamics,maitre2020dark}.

\begin{acknowledgments}
This work was supported by the Russian Science Foundation project N\textsuperscript{\underline{\scriptsize o}} 25-72-31023.
\end{acknowledgments}

\bibliographystyle{apsrev4-1}

\end{document}